\documentclass[letterpaper]{article} 
\usepackage{aaai24}  
\usepackage{times}  
\usepackage{helvet}  
\usepackage{courier}  
\usepackage[hyphens]{url}  
\usepackage{graphicx} 
\usepackage{natbib}  
\usepackage{caption} 
\usepackage{booktabs}
\usepackage{multirow}

\title{Deep Learning-based Prediction of Breast Cancer Tumor and Immune Phenotypes from Histopathology}
\author {
    Tiago Gonçalves\textsuperscript{\rm 1,\rm 2, \rm 3}, 
    Dagoberto Pulido-Arias\textsuperscript{\rm 1}, 
    Julian Willett\textsuperscript{\rm 1}, 
    Katharina V. Hoebel\textsuperscript{\rm 1}, 
    Mason Cleveland\textsuperscript{\rm 1}, 
    Syed Rakin Ahmed\textsuperscript{\rm 1}, 
    Elizabeth Gerstner\textsuperscript{\rm 1}, 
    Jayashree Kalpathy-Cramer\textsuperscript{\rm 1}, 
    Jaime S. Cardoso\textsuperscript{\rm 2, \rm 3}, 
    Christopher P. Bridge\textsuperscript{\rm 1}\equalcontrib, 
    Albert E. Kim\textsuperscript{\rm 1}\equalcontrib
}
\affiliations {
    \textsuperscript{\rm 1}Athinoula A. Martinos Center for Biomedical Imaging, Massachusetts General Hospital, Harvard Medical School\\
    \textsuperscript{\rm 2}Faculty of Engineering, University of Porto\\
    \textsuperscript{\rm 3}Institute for Systems and Computer Engineering, Technology and Science (INESC TEC)\\
    Corresponding author: akim46@mgh.harvard.edu
}

\begin{document}

\maketitle

\begin{abstract}
The interactions between tumor cells and the tumor microenvironment (TME) dictate therapeutic efficacy of radiation and many systemic therapies in breast cancer. However, to date, there is not a widely available method to reproducibly measure tumor and immune phenotypes for each patient’s tumor. Given this unmet clinical need, we applied multiple instance learning (MIL) algorithms to assess activity of ten biologically relevant pathways from the hematoxylin and eosin (H\&E) slide of primary breast tumors. We employed different feature extraction approaches and state-of-the-art model architectures. Using binary classification, our models attained area under the receiver operating characteristic (AUROC) scores above $0.70$ for nearly all gene expression pathways and on some cases, exceeded 0.80.  Attention maps suggest that our trained models recognize biologically relevant spatial patterns of cell sub-populations from H\&E.  These efforts represent a first step towards developing computational H\&E biomarkers that reflect facets of the TME and hold promise for augmenting precision oncology.
\end{abstract}

\section{Introduction}\label{sec:intro}
A cardinal facet of precision oncology is using complementary biological facets from a patient’s tumor to optimize clinical decision making.  While genomic sequencing of tumor tissue has transformed the therapeutic landscape of modern oncology, it has become increasingly apparent that genomic biomarkers (DNA sequencing analysis) are insufficient as stand-alone tools, given the wide variability in clinical outcomes for patients with similar biomarker profiles. To improve upon this limitation, some have proposed incorporating transcriptional profiling (RNA sequencing analysis) to the precision oncology framework to define improved biomarkers of efficacy~\cite{lin2019pros}. Recent studies have linked immune activity, angiogenesis, and tumor mitoses pathways in the TME to risk of metastasis and therapeutic efficacy~\cite{khan2018improving}, thus highlighting a need to monitor therapeutically relevant facets of the TME. 

To date, there is not a commercially available method in clinical practice to evaluate a patient’s TME in clinical oncology. Efforts to develop a widely available gene expression panel quantifying facets of the TME have been limited due to the need for RNA extraction from fresh tissue instead of the more widely available paraffin-embedded tissue. Furthermore, pathology workflows mainly evaluate tumor cell characteristics (e.g., grading, subtyping) within H\&E and are not able to reproducibly assess the TME. As spatial profiling efforts suggest that the physical colocalization of non-malignant cells to tumor cells modulate intracellular communication, tumor phenotypes, and clinical outcomes~\cite{longo2021integrating}, there is likely biologically relevant information from H\&E that is not used in clinical care. Therefore, given the importance of the TME and wide availability of H\&E, we sought to build upon recent advances in computer vision to estimate tumor and immune phenotypes from H\&E whole slide images (WSI).

\begin{figure*}
    \centering
    \includegraphics[width=0.80\linewidth]{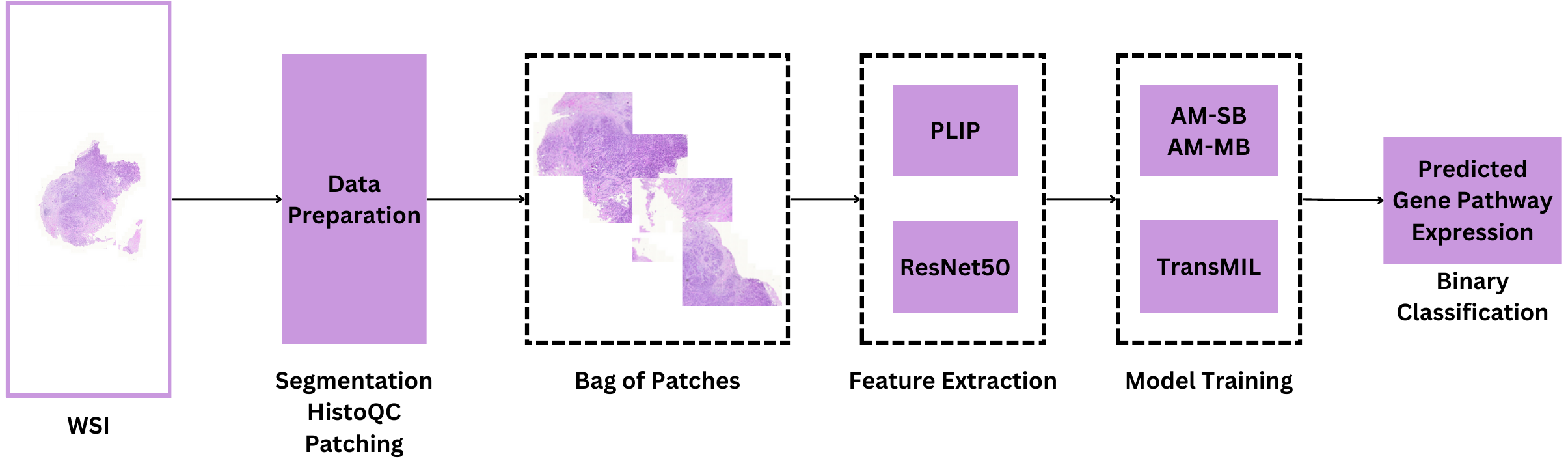}
    \caption{Overview of the proposed pipeline.}
    \label{fig:graphical_abstract}
\end{figure*}

Here, we present an exploratory application of MIL algorithms to characterize tumor and immune phenotypes from the H\&E slide of primary breast tumors.  We hypothesized that deep learning (DL) could predict these facets of the TME through recognizing biologically relevant features (e.g., tissue architecture, tumor-immune interface) within an individual H\&E image.  Using a DL classification pipeline, we employed different feature extraction techniques and model architectures to optimize the prediction of tumor and immune phenotypes (see Figure~\ref{fig:graphical_abstract}).  To the best of our knowledge, our effort is one of the first to comprehensively monitor the TME from H\&E using an end-to-end DL approach. This work can inspire collaboration for computational pathology biomarker development to facilitate development of a widely available tool that quantifies facets of the TME for each patient. The source code and implementation details are publicly available in a GitHub repository\footnote{\url{https://github.com/QTIM-Lab/dl-prediction-brca-tiph}}.

\section{Related Work}\label{sec:related-work}
With the increasing availability of large biobanks of patient-derived H\&E samples, multi-Omics, and high-performance computing capabilities, there is great promise to combine these tools towards driving new insights in oncology~\cite{montezuma2023,app10144728,cancers14102489,wu2023100184}. However, analysis of H\&E WSIs is computationally difficult due to their multi-pyramidal gigapixel resolution. Down-sampling these images is undesirable, as this would result in a loss of biologically-relevant information that may harm a DL model’s performance. One popular strategy for WSI-based DL analysis is weakly-supervised MIL, as this strategy avoids the need for manual data annotations. After a WSI is cropped into patches, feature vector embeddings are extracted from each patch individually.  The collection of patch-level feature vectors is then inputted into a MIL framework to yield the corresponding WSI-level representation and score~\cite{srinidhi2021}.

One example of this strategy is the Clustering-constrained Attention Multiple instance learning approach (CLAM)~\cite{lu2021data}. This method uses attention-based pooling to aggregate patch-level representations, weighted by the attention score, into slide-level representations.  As this slide-level representation, instead of the entire WSI, is the input to the final classification module, CLAM requires fewer computational resources to train and optimize.  However, an important limitation of MIL is that each patch is assumed to be independent of surrounding patches~\cite{lipkova20221095}. This assumption may not be valid, given the importance of spatial localization of different cell populations in dictating tumor phenotype.  To this end, recently developed vision transformer models account for spatial correlation and context between patches. The Transformer based MIL (TransMIL)~\cite{shao2021transmil} and the Hierarchical Image Pyramid Transformer (HIPT)~\cite{chen2022self,chen2022scaling} leverage morphological and spatial information to learn representations that encode the natural hierarchical structure inherent in WSIs. While there is early evidence that these architectures may boost performance for tasks in computational pathology, it is not immediately clear whether this boost in performance may justify the increased computational resources that vision transformers require.  Therefore, we evaluated the impact of the architecture on model performance in predicting tumor and immune phenotypes. 

Another timely question for H\&E-based model optimization is the best strategy for feature extraction from the WSI.  While non-medical DL models pre-trained on ImageNet have achieved tremendous performance, it is unclear whether this strategy is optimal for H\&E-based tasks. A prevailing hypothesis is that models pre-trained on H\&E data would have improved performance on H\&E-based tasks, compared to a model pre-trained on real-world images~\cite{lu2023cvpr,huang2023visual}.  To this end, recent efforts have combined large language models and image analysis pipelines with self-supervised learning to create H\&E-based foundation models. The MI-Zero framework~\cite{lu2023cvpr} and pathology language–image pre-training (PLIP) foundation model~\cite{huang2023visual} use visual-language encoders to extract image and text features, compute the similarity scores between text and image embeddings, and employs an aggregation function to generate a slide-level prediction. Here, we evaluated the optimal method for feature extraction by applying both an ImageNet- and an H\&E-based feature extractor to predict TME phenotypes.

\begin{table*}[htbp]
\centering
\caption{\textbf{AUROC results on the test set, obtained for all model architectures, using ResNet50- and PLIP-derived features.} The best results per type of feature-extraction strategy are highlighted in bold. While the AM-SB and AM-MB model architectures outperform TransMIL prediction of all gene expression tasks, the increase in performance was marginal, suggesting that model architecture may not play a critical role in the classification of gene expression.}
\label{tab:all_models}
\begin{tabular}{@{}lcccc@{}}
\toprule
\multicolumn{1}{c}{\multirow{3}{*}{\textbf{Task}}} & \multirow{3}{*}{\textbf{Class Proportion (0 / 1)}} & \multicolumn{3}{c}{\textbf{Architecture \& Features}} \\ \cmidrule(l){3-5} 
\multicolumn{1}{c}{} &  & AM-SB & AM-MB & TransMIL \\
\multicolumn{1}{c}{} &  & ResNet50 / PLIP & ResNet50 / PLIP & ResNet50 / PLIP \\ \cmidrule(r){1-2}
B-cell proliferation & 60.45\% / 39.55\% & 0.6974 / \textbf{0.7755} & \textbf{0.7322} / 0.7735 & 0.6960 / 0.7673 \\
T-cell mediated cytotoxicity & 53.72\% / 46.28\% & 0.7149 / 0.7703 & \textbf{0.7564} / \textbf{0.7770} & 0.7223 / 0.7196 \\
Angiogenesis & 47.92\% / 52.08\% & 0.7042 / \textbf{0.7435} & \textbf{0.7053} / 0.7213 & 0.7019 / 0.7214 \\
Epithelial-mesenchymal transition & 44.08\% / 55.92\% & \textbf{0.8110} / 0.7848 & 0.8023 / \textbf{0.8082} & 0.7545 / 0.7934 \\
Fatty acid metabolism & 52.83\% / 47.17\% & \textbf{0.6323} / \textbf{0.6030} & 0.6294 / 0.5920 & 0.5258 / 0.5634 \\
Glycolysis & 49.31\% / 50.69\% & \textbf{0.7996} / 0.8118 & 0.7954 / \textbf{0.8330} & 0.7834 / 0.8045 \\
Oxidative phosphorylation & 49.50\% / 50.50\% & 0.6894 / 0.7145 & \textbf{0.6926} / \textbf{0.7332} & 0.6699 / 0.6826 \\
Immunosuppression & 59.51\% / 40.49\% & 0.7996 / 0.8458 & \textbf{0.8133} / \textbf{0.8542} & 0.7572 / 0.8113 \\
Antigen processing and presentation & 49.50\% / 50.50\% & 0.7450 / 0.7806 & \textbf{0.7599} / \textbf{0.7924} & 0.7503 / 0.7342 \\
Cell cycling & 51.26\% / 48.74\% & 0.7768 / \textbf{0.7939} & \textbf{0.7809} / 0.7852 & 0.7121 / 0.7229 \\ \bottomrule
\end{tabular}
\end{table*}

\section{Methodology}\label{sec:methodology}
\subsection{WSI Data Preparation}
We used the breast invasive carcinoma (BRCA) subset of The Cancer Genome Atlas (TCGA): TCGA-BRCA~\cite{tcgabrca2016}. For quality assessment, we used the default parameters of HistoQC~\cite{histoqc2019,histoqc2021}.  We obtained a curated segmentation mask (i.e., tissue \textit{versus} background) for each WSI. Next, using this mask, we split each WSI ($20\times$, $40\times$ magnification) into patches of size $256\times256$ using the pipeline described in~\cite{lu2021data}. Finally, we extracted feature vectors from each patch using
two different approaches: a ResNet50~\cite{he2016deep,lu2021data} backbone pre-trained on the ImageNet dataset~\cite{ILSVRC15} and the PLIP~\cite{huang2023visual} image encoder pre-trained on the OpenPath dataset~\cite{huang2023visual}.

\subsection{Quantification of Gene Pathway Expression}
We obtained the normalized expression of individual genes (bulk RNA-sequencing data) for the TCGA-BRCA cohort from the cBioPortal for Cancer Genomics~\cite{cbioportal2023}. Each biological pathway was defined as per Gene Set Enrichment Analysis (GSEA) protocols. We then employed single-sample GSEA (ssGSEA) to calculate gene set enrichment scores for each sample and pathway pairing. ssGSEA transforms a single sample's gene expression profile to a gene set enrichment profile. This enrichment score represents the activity level of the biological pathway in which the gene set's members are coordinately up (i.e., positive values) or down-regulated (i.e., negative values)~\cite{subramanian2005gene,coscia2018multi,yi2020ssgsea}. 

\subsection{Models and Training Details}
We approached the problem as a binary classification task. To this end, the ssGSEA scores underwent binarization, with negative scores receiving the label $0$ and the rest label $1$. A random data split allocated $70\%$ to training, $15\%$ to validation, and $15\%$ to testing, ensuring exclusive assignment of a given patient's data to one split. We trained the models using CLAM~\cite{lu2021data} and TransMIL~\cite{shao2021transmil}.

\subsubsection{CLAM}
There are two variations of this architecture: single-branch CLAM (CLAM-SB), which employs a fully connected layer for slide-level classification, and multi-branch CLAM (CLAM-MB), which uses a fully connected layer per class, using the class with the maximum score for slide-level classification. CLAM also promotes instance-level clustering under a mutual exclusivity assumption (e.g., cancer subtypes are mutually exclusive: when one subtype is present in the WSI, we may assume that there is no morphology of the other subtypes concurrently present).
Since this assumption does not apply to our case, we only used the attention mechanism (along with the classifier), giving us two model architectures: AM-SB and AM-MB. We trained these models for 300 epochs, using cross-entropy as the loss function and the adaptive moment estimation (Adam) optimizer~\cite{kingma2014adam} with a learning rate of $2\times10^{-4}$ and weight decay of $1\times10^{-5}$. We saved the best model according to the loss on the validation set.

\subsubsection{TransMIL}
We trained TransMIL for 200 epochs, using cross-entropy as the loss function and the Adam optimizer with a learning rate of $2\times10^{-4}$ and weight decay of $1\times10^{-5}$. We saved the best model according to the loss on the validation set.

\section{Results and Discussion}\label{sec:results-discussion}
Consistent with our hypothesis, our models predict TME phenotypes from H\&E WSIs to reasonable accuracy. Table~\ref{tab:all_models} presents the area under the receiver operating characteristic curve (AUROC) results on the test for both model architectures, using either ResNet50-derived or PLIP-derived features.  First, we observe that models predicting immune phenotypes (e.g., B-cell proliferation, T-cell mediated cytotoxicity, immunosuppression, antigen presentation) achieve satisfactory performance (AUROC 0.75-0.8). Given prior work linking the spatial distribution of tumor-infiltrating lymphocytes and myeloid cells to immune activity~\cite{longo2021integrating}, our findings illustrate DL’s ability to identify and learn spatial patterns of cell populations predictive of specific facets of immune function. These findings are noteworthy, as the type (e.g., T-, B-, and dendritic cell) or function (e.g., cytotoxic or exhausted) of lymphocyte is indistinguishable from manual pathology review on H\&E. Future work is needed to identify specific spatial patterns on H\&E indicative of immune activity or exhaustion.

Next, we observed that our models predicting facets of tumor aggressiveness (e.g., cell cycling, angiogenesis, and epithelial-mesenchymal transition (EMT)) achieve reasonable performance and can be estimated from H\&E. These findings are biologically plausible, as cell cycling measures mitotic activity and aggressiveness of tumor cells, angiogenesis represents growth of new blood vessels (a hallmark of cancer growth), and EMT measures a tumor's ability to invade surrounding tissue.  Biological activity associated with these pathways can be observed on H\&E, but are often difficult for pathologists to quantify. Conversely, we observe fatty acid metabolism and oxidative phosphorylation achieving relatively poor performance. These tasks are metabolic pathways that are often not apparent in cell morphology and tissue architecture on H\&E and likely require more specific stains to measure.

From a technical perspective, we observe AM-SB and AM-MB architectures marginally outperform TransMIL-based prediction of all gene expression tasks, suggesting that the type of architecture likely does not play a crucial role in model performance. On the other hand, models trained with PLIP-based features generally outperform those trained with ResNet50-based features (see Table~\ref{tab:best_models_fts}), suggesting that using a feature extractor pre-trained on H\&E-related data allows the model to better learn essential features or relationships specific to histopathology. Importantly, PLIP-based features (512-dimensional feature vectors) also possess a computational advantage since they are half the size of ResNet50 features (1024-dimensional feature vectors), thus being more manageable to store and process.

\begin{figure}[htbp]
    \centering
    \includegraphics[width=\linewidth]{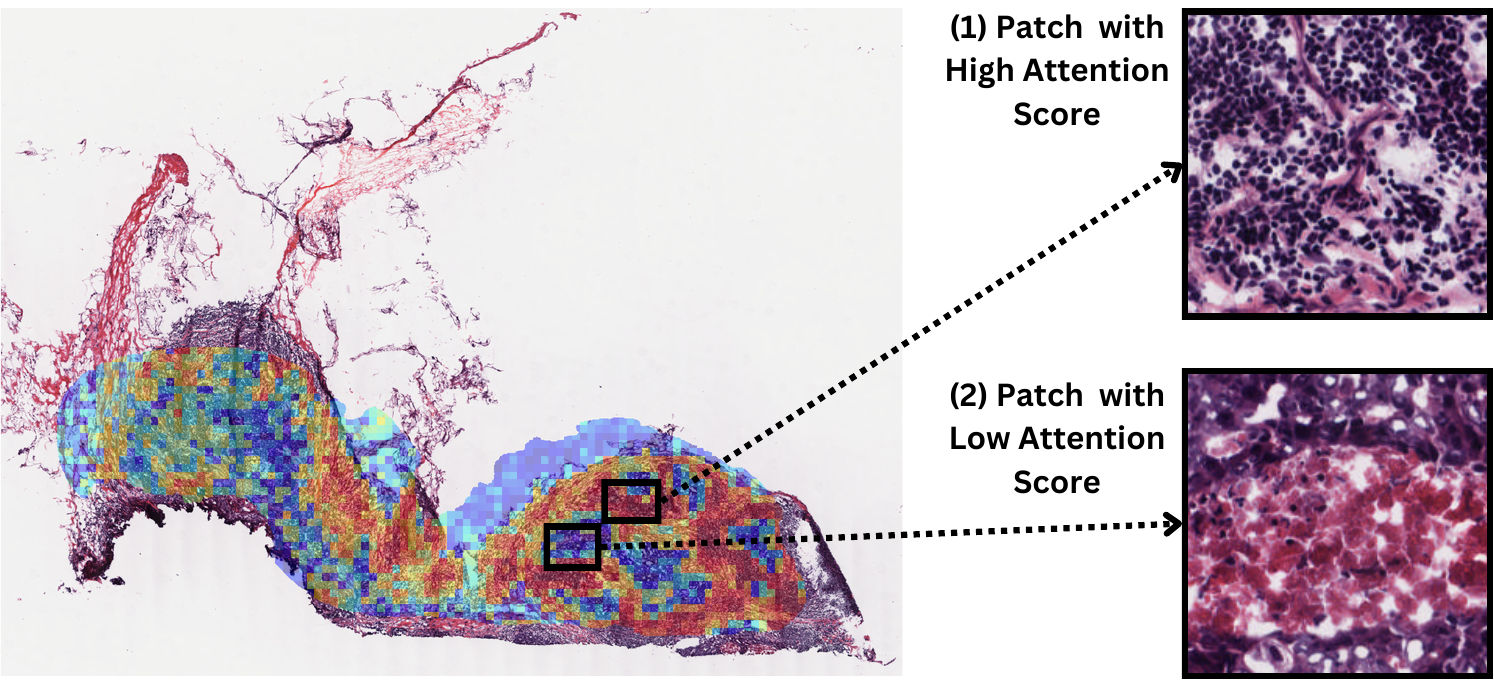}
    \caption{\textbf{Attention map obtained using the AM-SB architecture for a WSI predicted to have a high-degree of T-cell mediated cytotoxicity}. Red zones and blue zones indicate high and low attention scores, respectively. On the right, we provide two exemplar patches: 1) the high-attention patch illustrates abundant tumor-infiltrating lymphocytes without tumor cells, which are suggestive of high immune activity, and 2) the low-attention region demonstrated areas of tumor necrosis and minimal lymphocytes, consistent with low immune activity.}
    \label{fig:att-maps}
\end{figure}

Finally, we sought to identify explainable morphological descriptors that may govern tumor and immune function from our trained models.  We generated an attention map on an H\&E WSI that was predicted to have a high level of T-cell mediated cytotoxicity (see Figure~\ref{fig:att-maps}).  We note that a high-attention patch corresponded with a high degree of infiltrating lymphocytes and a low-attention patch highlighted areas of necrotic tumor. These attention maps provide additional evidence, albeit preliminary, that our trained models are learning therapeutically-relevant patterns. Further work is needed to rigorously evaluate and optimize this approach.

\begin{table}[htbp]
\centering
\caption{\textbf{AUROC results on the test set, obtained for the best model of each feature set.} The best-performing feature-extraction strategy is highlighted in bold. Models trained with PLIP-derived features generally perform better than ResNet-derived features, suggesting that using a feature extractor pre-trained on H\&E-related data may improve performance.}
\label{tab:best_models_fts}
\begin{tabular}{@{}lcc@{}}
\toprule
\multicolumn{1}{c}{\multirow{3}{*}{\textbf{Task}}} & \multicolumn{2}{c}{\textbf{Features}} \\ \cmidrule(l){2-3} 
\multicolumn{1}{c}{} & ResNet50 & PLIP \\ \cmidrule(r){1-1}
B-cell proliferation & 0.7322  & \textbf{0.7755} \\

T-cell mediated cytotoxicity & 0.7564 & \textbf{0.7770} \\

Angiogenesis & 0.7053 & \textbf{0.7435} \\

Epithelial-mesenchymal transition & \textbf{0.8110} & 0.8082 \\

Fatty acid metabolism & \textbf{0.6323} & 0.6030 \\

Glycolysis & 0.7996 & \textbf{0.8330} \\

Oxidative phosphorylation & 0.6926 & \textbf{0.7332} \\

Immunosuppression & 0.8133 & \textbf{0.8542} \\

Antigen processing and presentation & 0.7599 & \textbf{0.7924} \\

Cell cycling & 0.7809 & \textbf{0.7939} \\ \bottomrule
\end{tabular}
\end{table}

\section{Conclusion and Future Work}\label{sec:conclusion-fw}
Our study represents one of the first efforts to predict therapeutically relevant facets of the TME from an H\&E WSI.  Here, our work suggests that tumor and immune phenotypes can be inferred with reasonable accuracy using state-of-the-art DL methods. Importantly, we note that the choice of feature extraction strategy may impact performance.  A model pre-trained on H\&E-based features had improved performance for most tasks, compared to a model pre-trained on ImageNet-based features. While additional work is needed to build upon and validate this work, we believe our results may be of interest. First, for several other medical tasks, feature extraction strategies more native to the modality of interest do not always outperform natural image-based feature extraction~\cite{Kornblith_2019_CVPR,Zhou2020,morid2021scoping,li2022improving}. Second, as prediction of tumor and immune phenotypes can be inferred from H\&E, these computational H\&E biomarkers would have major implications for cancer patients and precision oncology as many hospitals only possess the capability for H\&E and lack the resources for DNA/RNA sequencing.

In future work, we will experiment with other feature extraction frameworks (e.g., HIPT, MI-Zero) and feature fusion approaches (e.g., combining two different feature sets in a single model). Additionally, we intend to explore ordinal and multi-task classification and measures to boost model interpretability. Finally, we will trial multi-modal architectures, integrating relevant clinical metadata during training to assess the impact on performance.

\section{Acknowledgments}\label{sec:ackn}
The results published or shown here are in whole or part based upon data generated by the TCGA Research Network\footnote{\url{https://www.cancer.gov/tcga}}. This work was funded by the William G. Kaelin, Jr., M.D., Physician-Scientist Award of the Damon Runyon Cancer Research Foundation, American Association for Cancer Research Breast Cancer Research Fellowship, and Conquer Cancer/American Society of Clinical Oncology Young Investigator Award, and by the Portuguese Foundation for Science and Technology (FCT) through the Ph.D. grant ``2020.06434.BD''.

\bibliography{aaai24}

\end{document}